\documentclass[reprint,amsmath,amssymb,aps,showkeys,nofootinbib]{revtex4-2}

\usepackage{amsmath,amssymb}
\usepackage{graphicx}
\usepackage{xcolor}
\usepackage{hyperref}
\hypersetup{hidelinks,pdftitle={Dressing Away the Shift in BF Theory},pdfauthor={J. Francois and L. Ravera}}
\usepackage{upgreek}
\usepackage{mathrsfs}
\DeclareFontFamily{OT1}{pzc}{}
\DeclareFontShape{OT1}{pzc}{m}{it}{<-> s * [1.1] pzcmi7t}{}
\DeclareMathAlphabet{\mathpzc}{OT1}{pzc}{m}{it}

\newcommand{\Lie}{\mathrm{Lie}}
\newcommand{\dd}{\mathrm{d}}

\newcommand{\hh}{\mathcal{H}}

\newcommand{\vphi}{\varphi}

\renewcommand\-{^{-1}}

\newcommand{\Tr}{\operatorname{Tr}}

\newcommand{\cB}{\mathcal{B}}
\newcommand{\cT}{\mathcal{T}}
\newcommand{\cK}{\mathcal{K}}
\newcommand{\cJ}{\mathcal{J}}

\begin{document}

\title{Higher Dressing and Shift-Invariance in 4D Topological BF Theory}

\author{J. Fran\c{c}ois}
\email{jordan.francois@uni-graz.at}
\affiliation{University of Graz (Uni Graz), Heinrichstra\ss e 26/5, 8010 Graz, Austria, and \\ Masaryk University (MUNI), Kotl\'{a}\v{r}sk\'{a} 267/2, Veve\v{r}\'{i}, Brno, Czech Republic, and \\ Mons University (UMONS), 20 Place du Parc, 7000 Mons, Belgium.}
\author{L. Ravera}
\email{lucrezia.ravera@polito.it}
\affiliation{Politecnico di Torino (PoliTo), C.so Duca degli Abruzzi 24, 10129 Torino, Italy, and \\ Istituto Nazionale di Fisica Nucleare (INFN), Section of Torino, Via P. Giuria 1, 10125 Torino, Italy, and \\ Grupo de Investigaci\'{o}n en F\'{i}sica Te\'{o}rica (GIFT), Universidad Cat\'{o}lica De La Sant\'{i}sima Concepci\'{o}n, Alonso de Ribera 2850, Concepci\'{o}n, Chile.}

\date{\today}

\begin{abstract}
A characteristic feature of topological field theories is the absence of local propagating degrees of freedom. In four-dimensional BF theory with cosmological term, we make this feature manifest by exactly reducing the finite non-Abelian shift symmetry. 
We do so by identifying a \emph{higher dressing field}, a 1-form compensating the shift, allowing to build from first principles the composite shift-invariant field $\cB=B-\lambda^{-1}F$. In terms of this dressed field, the dynamics becomes algebraic and sets $\cB=0$, directly exposing the absence of local propagating degrees of freedom. 
This provides the first explicit instance of the higher-form Dressing Field Method.
\end{abstract}

\keywords{Topological field theory, BF theory, Shift symmetry, Dressing Field Method, Higher dressing.}
\maketitle

\section{Introduction}
\label{Introduction}

In general-relativistic gauge field theory (gRGFT), local symmetries identify field configurations that describe the same physical state. 
Extracting invariant variables makes this explicit and can reveal a theory's 
physical degrees of freedom (d.o.f.)
before  quantization. 
The Dressing Field Method (DFM) is a systematic mean to obtain such variables, via field-dependent \emph{dressing fields}
\cite{JTF-Ravera2024gRGFT}. 
The DFM is technically and conceptually distinct from gauge fixing \cite{Francois-Berghofer2024,JTF-Ravera2025-MFSRelaionalQuant}.

Applied to (classical and quantum \cite{JTF-Ravera-InvQuantLetter2026}) gRGFT,  it provides variables invariant under both gauge and diffeomorphism transformations. 
It is background independent, intrinsically non-perturbative, yet admits a perturbative implementation, and naturally enjoys a \emph{relational} interpretation \cite{JTF-Ravera2024gRGFT,JTF-Ravera-Foundation2025}.
Its applies beyond ``conventional" internal gauge transformations, as illustrated by recent work on supersymmetric field theory \cite{JTF-Ravera2024-SUSY,JTF-Ravera-UnconventionalSUSY,JTF-Ravera2025offshellsusy}.

The question addressed here is whether the same approach can be applied to  transformations parametrized by differential forms, yielding \emph{higher symmetries}, thus providing a first explicit example of \emph{higher dressing}. 
In this context, BF theories are a  class of topological
field theories whose symmetry structure includes \emph{shifts}, i.e., transformations parametrized by differential forms
\cite{Blau:1989bq}.

Moreover, since topological BF theories have no local propagating d.o.f., reducing their shift symmetry offers a natural way to examine how this absence manifests itself in the reduced description: the reduction should make this absence more transparent.

We shall thus consider four-dimensional BF theory with cosmological term as a test case, which furthermore displays an interesting interplay between ordinary gauge transformations
and shift symmetry
\cite{Baez:1995ph,Cattaneo:1997eh,Freidel:2005ak,
deGracia:2017ejv,Cattaneo:2025xvk}.
This theory features a connection $A$ and an adjoint-valued 2-form $B$ in a metric-independent (hence topological) action containing the quadratic coupling $-\tfrac{\lambda}{2}\Tr(B\wedge B)$, with $\lambda\ne0$, called the ``cosmological term'' -- though $\lambda$ is not meant to be the physical cosmological constant.  
Its quantization yields a topological quantum field theory \cite{Baez:1995ph}. 
Its higher-gauge interpretation also makes it a natural testing ground for symmetry reduction beyond ordinary gauge groups \cite{Baez-Huerta2011}. 
The model admits two types of local transformations. 
On the one hand, ordinary gauge transformations, parametrized locally by group-valued 0-forms, change the internal d.o.f.: 
$A$ transforms as a connection and $B$ by conjugation. 
On the other hand, \emph{shifts} are parametrized by adjoint-valued 1-forms. 

We show that the connection $A$ induces an example of  higher dressing:
We define the shift-dressing $u[A]=-A/\lambda$, obeying $u^\eta=u-\eta$, thus compensating the shift of the original fields. 
It allows to build a shift-invariant dressed 2-form $\cB=B-\lambda^{-1}F$. 
The ordinary gauge symmetry remains, acting by conjugation on $\cB$ (a.k.a., in the DFM literature, residual transformations of the 1st kind \cite{JTF-Ravera2024gRGFT}).
The dressed bulk equation becomes $\cB=0$, displaying the absence of local bulk degrees of freedom. 

This Letter is organized as follows. 
In Section \ref{DFM}, we briefly recall the DFM, showing how it applies to additive higher-form symmetries. 
In Section \ref{BF}, we describe the cosmological BF model and its finite non-Abelian shift symmetry, as well as the latter's relation to ordinary gauge transformations of the model. 
In Section \ref{Construction}, 
we treat the 4D BF model via the DFM, deriving the dressed bulk dynamics. 
Finally, in Section \ref{Discussion}, we discuss subsequent developments downstream of the present work.

\section{The Dressing Field Method: Basics}
\label{DFM}

Let $\Phi$ denote the \emph{field space} of a gRGFT, with points $\vphi$ and a right action $\vphi\mapsto\vphi^\gamma$ of a gauge group $\hh$. 
For a subgroup $\cK\subseteq\hh$, a \emph{field-dependent dressing field} obeys
\begin{align}
u[\vphi]^\kappa := u[\vphi^\kappa]=\kappa^{-1}u[\vphi],\quad \kappa\in\cK.
 \label{DFMgaugetr}
\end{align}
Note that the transformation law distinguishes a dressing from a gauge transformation \cite{Francois-Berghofer2024,JTF-Ravera2024gRGFT};
therefore, \eqref{DFMgaugetr} is not merely a field-dependent gauge transformation. 
One then defines the \emph{dressed fields}
\begin{align}
\label{dressed-fields-generic}
\vphi^u:=\vphi^{u[\vphi]}, 
\end{align}
automatically $\cK$-invariant, $(\vphi^u)^\kappa :=(\vphi^\kappa)^{u[\vphi]^\kappa}=\vphi^{u[\vphi]}$. 
Remark that dressed fields are not gauge-fixed variables.
For $\vphi=\{A, \psi\}$, with $A$ a connection and $\psi$ a matter field in a representation $\rho$, \eqref{dressed-fields-generic} yields the  dressed variables 
\begin{equation}
\begin{aligned}
 A^u&=u^{-1}Au+u^{-1}\dd u, \\
 \psi^u&=\rho(u)^{-1}\psi.
 \label{usualdressing}
\end{aligned}
\end{equation}
The $\cK$-invariant dressed fields \eqref{dressed-fields-generic} are \emph{relational} variables: Since $u=u[\vphi]$, $\vphi^u$ represents the coordinatization by the d.o.f. of $\vphi$ entering $u$ of  $\vphi$'s remaining d.o.f., i.e., $\vphi^u$ encodes the $\cK$-\emph{invariant relations} between $\vphi$'s degrees of freedom.

If $\cK$ is a normal subgroup, $\cK \triangleleft \hh$, so that \mbox{$\hh = \cK \ltimes \cJ$}, then $\cJ=\hh/\cK$ acts as a group of residual gauge transformations (a.k.a. \emph{residual transformations of the 1st kind}) of the dressed variables \eqref{dressed-fields-generic}-\eqref{usualdressing},
determined by the action of $\cJ$ on $\vphi$ (known \emph{a priori}) and $u=u[\vphi]$ (through $\vphi$).

\smallskip

Consider an \emph{additive action} of the Abelian group $\cT=\Omega^\bullet(M)$ on $\Phi$ defined as  $T_\eta \vphi := \vphi +\eta =: \vphi^\eta$, with $\eta=\eta_I$ and $I=\{0, 1, \ldots\}$ 
the multi-index indicating the differential form degrees of the set of fields $\vphi=\vphi_I=\{\vphi_0, \vphi_1, \ldots\}$. 
One allows for the possibility that, for some degree $i$, $\eta_{i+1}=p(\eta_i, \vphi_i)$ for $p$ a polynomial function. 
The group $\cT$ being additive Abelian, its action on $\Phi$ is
$T_{\eta_2}T_{\eta_1}=T_{\eta_1+\eta_2}=T_{\eta_1}T_{\eta_2}$, for $\eta_1, \eta_2 \in \cT$.

\smallskip

Adapting the Abelian version of \eqref{DFMgaugetr}, a $\vphi$-dependent $\cT$-dressing field
$u_I=u[\vphi] \in \Omega^I(M)$ satisfies
\begin{align}
T_\eta u[\vphi]:=
 u[T_\eta\vphi]=u[\vphi]-\eta.
 \label{additivedressing}
\end{align}
The $\cT$-dressed fields are defined by
\begin{align}
\label{T-dressed fields}
\vphi^{u_I}:=T_{u[\vphi]}\vphi =\vphi + u_I,
\end{align}
and are $\cT$-invariant by construction: 
\begin{align*}
 (\vphi^{u_I})^\eta :=(T_\eta\vphi)^{u[T_\eta\vphi]}
 =T_{u[\vphi]-\eta}T_\eta\vphi =T_{u[\vphi]}\vphi
 =\vphi^{u_I}.
\end{align*}
Remark that this construction is finite, i.e., holds at the group level, not only at the linear, Lie algebraic, level.

\section{BF theory and its  shift symmetry}
\label{BF}

Let $P\to M$ be a principal $H$-bundle over an oriented four-manifold $M$, with ordinary (0-form) gauge group $\hh$. Let $\mathfrak h=\Lie H$ carry a nondegenerate invariant symmetric bilinear form, denoted $\Tr$. 
The field content of the BF theory we consider is given by $\vphi=\{A, B\}$ where $A \in \Omega^1(M, \mathfrak h)$ is a connection  with curvature $F:=dA + A \wedge A\, \in\Omega^2(M, \mathfrak h)$, and  $B\in\Omega^2(M,\mathfrak h)$.
Their $\hh$-transformations are, for $\gamma \in \hh$,
\begin{equation}
\begin{aligned}
A^\gamma&=\gamma^{-1}A\gamma+\gamma^{-1}\dd\gamma,
\quad 
F^\gamma=\gamma^{-1}F\gamma,
\\
B^\gamma&=\gamma^{-1}B\gamma. \label{ordinarygauge} 
\end{aligned}
\end{equation}
Let us define the 1-form shift group $\cT:=\Omega^1(M,\mathfrak h)$ acting, for $\eta \in \cT$, as
\begin{equation}
\begin{aligned}
 A^\eta&=A+\lambda\eta,
 \quad 
 F^\eta=F+\lambda D_A\eta+\lambda^2\eta\wedge\eta,\\
 B^\eta&=B+D_A\eta+\lambda\eta\wedge\eta,
 \label{finiteshift}
\end{aligned}
\end{equation}
for $\lambda\neq0$ a parameter, and where  
$D_A\eta:=\dd\eta+[A,\eta]$
with graded bracket, so that
$\tfrac12[\eta,\eta]=\eta\wedge\eta$. 
Meaning we have $\eta_I=\{\eta_1, \eta_2\}=\{\lambda \eta, D_A\eta + \lambda \eta\wedge \eta\}$, 
and $\vphi^{\eta_I}= \vphi+\eta_I$. 

The transformation $A^\eta$ is an automorphism of the space of connections iff $\cT$ supports the conjugation action by $\hh$, $\eta^\gamma=\gamma\- \eta \gamma$. 
Therefore, the local symmetry group is the semi-direct product group $\mathcal G = \hh \ltimes \cT$,  with $\cT \triangleleft \mathcal G$, and composition law (for its right action on $\Phi$):
\begin{align}
 (\gamma_1, \eta_1)*(\gamma_2 , \eta_2)
 :=
 \big(\gamma_1\gamma_2, \, 
 \eta_2 + \gamma_2^{-1} \eta_1 \gamma_2
 \big).
 \label{normality}
\end{align}
Its Lie algebra is $\mathfrak g=\mathfrak h \oplus\mathfrak t$, with Lie bracket
\begin{align}
 \big[ (\alpha_1, \zeta_1), \, (\alpha_2 , \zeta_2)
 \big]
 :=
 \big([\alpha_1, \alpha_2], \, 
 [\zeta_1, \alpha_2]-
 [\zeta_2, \alpha_1] 
 \big).
 \label{normality-alg}
\end{align}
It acts on the elementary fields $A$ and $B$ as
\begin{equation}
\begin{aligned}
\delta_{\alpha,\zeta}A&=D_A\alpha+\lambda\zeta,
\
 &&\delta_{\alpha,\zeta}B=[B,\alpha]+D_A\zeta  \\
 &=\delta_\alpha A + \delta_\zeta A ,
&&\phantom{\delta_{\alpha,\zeta}B}=\delta_\alpha B + \delta_\zeta B,
\label{infinitesimal}
\end{aligned}
\end{equation}
which are indeed the linearization of \eqref{finiteshift}.
Finally, one may observe (see \cite{Freidel:2005ak}) that  there is a $\cT$-invariant quantity
\begin{align}
 \cB:=B-\lambda^{-1}F, \quad \text{s.t.} \quad \cB^\eta=\cB,\quad \text{or}\quad
 \delta_\zeta \cB =0. 
 \label{composite}
\end{align}

The above pertains to the kinematics. 
The model is given by the action functional 
\begin{equation}
\begin{aligned}
S[A,B]
&=
\int_M L(A, B)  \\
&=
\int_M\Tr \left(B\wedge F-\frac\lambda2 B\wedge B\right),
 \label{action}
\end{aligned}
\end{equation}
that is $\hh$-invariant and quasi-invariant under $\cT$, 
\begin{align}
\label{GT-Lagrangian}
 L(A^\gamma, B^\gamma)&=L(A, B), \\[1mm]
  L(A^\eta, B^\eta)
  &=L(A, B) + d b(\eta; A), \notag\\
  &=L(A, B) 
   + \dd \Tr\Big(
\eta\, F
+\tfrac{\lambda}{2}\eta\, D_A\eta
+\tfrac{\lambda^2}{3}\eta^3
\Big), \notag
\end{align}
the wedge product of forms being tacit in the second line.
Consequently, the field equations
\begin{align}
 F-\lambda B=0,\quad D_AB=0, 
 \label{EOM}
\end{align}
are $\hh$-covariant and $\cT$-invariant. 
In particular one finds  
\begin{align}
D_{A^\eta} B^\eta = D_A B + \big[\eta, \lambda (B -\lambda^{-1}F) \big].
\end{align}
So the first equation ensures the second's $\cT$-invariance.
Note that the field equations \eqref{EOM} are not independent: the second  follows from the first and the Bianchi \mbox{identity} $D_A F=0$ which hold kinematically (i.e., identically).

\section{One-form dressing and dressed dynamics}
\label{Construction}

We now apply the DFM to  the cosmological BF model: Given the semi-direct structure of the local symmetry group $\mathcal G$, one ought to first try to reduce the shift group $\cT$ via a local 1-form dressing field,
thus leaving the  gauge group $\hh=\mathcal G / \cT$
as residual transformations.
Let us mention that this is similar to the DFM treatment of metric-affine gravity (MAG), whose \emph{a priori} local symmetry group $\mathcal G=\mathcal{GL}(n) \ltimes \cT^n$ is reduced to $\mathcal{GL}(n)$ through a (0-form) dressing  field $u[\xi]$ for the ``gauge translation" group $\cT^n$. Only, there $\xi$ is an \emph{ad hoc} field known as the radius vector in the  literature. 
The upshot of introducing an \emph{ad hoc} dressing is that MAG and Poincaré gravity are not of Yang-Mills-type, but actually reduce to standard Cartan geometric formulations of gauge gravity.
See \cite{JTF-Ravera-2025-MAG}.

\subsection{Dressed kinematics and residual symmetry}

By inspection of \eqref{finiteshift} it is easy to identity the $\cT$-dressing 1-form field 
$u[A]:=-\tfrac{1}{\lambda}A$ which satisfies the defining property \eqref{additivedressing},
\begin{align}
 u[A]^\eta :=
 u[A^\eta]=u[A]-\eta.
\end{align}
By \eqref{T-dressed fields}, one thus defines the $\cT$-invariant dressed fields $\vphi^{u_I}= \vphi + u_I$, and explicitly:
\begin{equation}
\label{dressed-fields}
\begin{aligned}
 A^u&=A+\lambda u=0,
 \\[1mm]
 B^u&=B+D_Au+\lambda u\wedge u
 \\
 &=B-\lambda^{-1}(\dd A+A\wedge A)=\cB.
\end{aligned}
\end{equation}
Here we used $D_AA=\dd A+2A\wedge A$. 
If follows that the dressed curvature  vanishes, $F^u=dA^u + A^u\wedge A^u \equiv0$, 
also checked via $F^u=F+\lambda D_A u+\lambda^2 u\wedge u = F -F=0$.
The DFM thus shows the  model's physical d.o.f. to be encoded by the $\cT$-invariant relational 2-form  $B^u\!=\!\cB$.  

As per the usual DFM philosophy, the residual $\hh$-transformation of the dressed fields \eqref{dressed-fields} depends on those of the bare fields $A$ and $B$, known by \eqref{ordinarygauge}, and of the dressing field $u=u[A]$ -- which, given its $A$-dependence, is also deduced from \eqref{ordinarygauge} -- and found to be
\begin{align}
 u[A]^\gamma
 =
 u[A^\gamma]
 =\gamma^{-1}u[A]\gamma
   -\lambda^{-1}\, \gamma^{-1}\dd\gamma.
\end{align}
From which follows that
\begin{align}
(A^u)^\gamma = \gamma^{-1} A^u \gamma =0, \quad \text{and}\quad 
\cB^\gamma=\gamma^{-1}\cB\gamma.
 \label{residual}
\end{align} 
The $\hh$-transformation of $\cB$ being also recovered from the $\hh$-transformations of the bare fields $B$ and $F$, by \eqref{composite},
\begin{align}
 \cB[A^\gamma,B^\gamma]
 =B^\gamma-\lambda^{-1}F^\gamma
 =\gamma^{-1}\cB[A,B]\gamma.
 \label{finiterestrdr}
\end{align}
Infinitesimally, this yields the $\mathfrak h$-action $\delta_\alpha\cB=[\cB,\alpha]$.
It is satisfying to obtain $\cB$ from first principles, as an instance of invariant dressed field built via the DFM.

\subsection{Dressed action and field equations}

The dressed Lagrangian is
\begin{equation}
\begin{aligned}
 L^u:=L(A^u, B^u)
 &=\Tr\left(
 B^u\wedge F^u-\frac{\lambda}{2}B^u\wedge B^u
 \right) \\[-.5mm]
 \hookrightarrow\ \ 
 L(\cB)
 &=-\frac{\lambda}{2}\Tr(\cB\wedge\cB).
 \label{dressedLagrangian}
\end{aligned}
\end{equation}
Contrary to the bare Lagrangian $L(A, B)$, which is $\cT$-quasi-invariant, $L^u=L(\cB)$ is strictly $\cT$-invariant, being a functional of the lone $\cT$-invariant dressed field $B^u=\cB$.
By \eqref{residual}, the dressed theory is clearly an $\hh$-gauge theory. 
The dressed Lagrangian and  the bare one belong to the same de Rham cohomology class; i.e., we have
\begin{equation}
\begin{aligned}
\label{dressed-L}
 L^u
  &=L + \dd b(u; A), \\
  &=L 
   + \dd \Tr\Big(
u\, F
+\tfrac{\lambda}{2}u\, D_A u
+\tfrac{\lambda^2}{3} u^3
\Big),
\end{aligned}
\end{equation}
as can be inferred from \eqref{GT-Lagrangian}.
Using the definition of the dressing $u=u[A]=-\tfrac{1}{\lambda} A$, it follows immediately that
\begin{align}
b(u; A)=-\tfrac{1}{2\lambda} \Tr\big( AF -\tfrac{1}{3} A^3\big)=:-CS_3(A; F)
\end{align}
is the 3D Chern-Simons functional.
Observe furthermore that  the automatic $\cT$-invariance of $L^u$ implies  
\begin{align}
\label{WZ-term-cocyclic-dressing}
b(u; A)^\eta = b(u; A) - b(\eta; A).
\end{align}
In the terms of \cite{JTF-Ravera2024gRGFT,JTF-Ravera2025-MFSRelaionalQuant,
JTF-Ravera-InvQuantLetter2026},  $b(u; A)$  is a \emph{cocyclic dressing} -- i.e., a (\emph{non ad hoc}, since $u$ is built from the field content of the theory) Wess-Zumino counterterm -- cancelling the classical $\cT$-anomaly $b(\eta; A)$ of the bare Lagrangian $L$. 
 
Now, eq. \eqref{dressed-L} means that the dressed and bare field equations are \emph{formally} ``the same", and encode the same bulk dynamics. 
Indeed, applying the variational principle to the dressed action
\begin{equation}
\label{dressedaction-1}
\begin{aligned}
S^u&=S[A^u, B^u] = \int_M L(A^u, B^u)\\[-.5mm]
&= S[\cB]
  =\int_M L(\cB ),
\end{aligned}
\end{equation}
we get, from the first form, the dressed field equations
\begin{equation}
\begin{aligned}
\label{dressed-FE-1}
 F^u-\lambda B^u&=0,
 \qquad\text{and}\
 &&D_{A}^u B^u=0, \\[-.5mm]
 -\lambda \cB&=0,
 && \ \ \ \ \dd \cB=0 ,
\end{aligned}
\end{equation}
where obviously the second equation follows from the first (as in the bare case),
or, from the second form, directly
\begin{align}
\label{dressed-FE-2}
-\lambda \cB=0.
\end{align}
Either way, by \eqref{dressed-fields}, the dressed field equation manifestly unfolds in terms of bare fields as $F- \lambda B=0$, reproducing the bare field equation \eqref{EOM}, 
proving that the dressed field equation encodes the same fundamental bulk dynamics. 
By
\eqref{residual}, the dressed field equation \eqref{dressed-FE-2} is $\hh$-covariant, 
as expected from the $\hh$-invariance of the dressed action.

\section{Discussion}
\label{Discussion}

The absence of local propagating d.o.f. 
is transparent in the shift-dressed cosmological BF theory, its dynamics being entirely expressed by an algebraic
equation for the $\cT$-invariant relational variable $B^u=\cB=0$.
Note that the absence of local bulk d.o.f. does not
exclude global topological information \cite{Baez:1995ph} or d.o.f. at a nontrivial boundary \cite{Cattaneo:2012qu}.
A detailed analysis of these aspects, however, lies beyond the scope of the present short communication and is left for future investigation.

From a broader perspective, this work is the first of its kind in extending the DFM's range range of \mbox{applicability}. 
The DFM is a systematic framework yielding dual invariant relational formulations of gRGFTs.
A broad range of invariant approaches in the field were shown to fall under its technical and conceptual purview -- with applications  in electroweak theory, gravity, twistor theory, conformal geometry, CFT, condensed matter physics,  supersymmetric field theory and supergravity, quantum reference frames, and even to classical point particles mechanics and non-relativistic quantum mechanics. 
See, e.g., \cite{JTF-Ravera2024gRGFT, JTF-Ravera2025-MFSRelaionalQuant} and references therein.

Here, for the first time, we show that the cosmological 4D BF theory provides an instance of higher DFM: 
its \emph{prima facie} higher local symmetry group $\mathcal G=\hh\ltimes \cT$ being reduced to a standard 0-form $\hh$ gauge group through a  dressing 1-form field $u=u[A]$ supplied by the connection $A$, and leaving its d.o.f. encoded by the relational $\cT$-invariant 2-form field $B^u=\cB$.

The quasi-$\cT$-invariance of the bare Lagrangian $L$, \eqref{GT-Lagrangian}, can be expressed as $L$ presenting a \emph{classical} ($d$-exact) higher $\cT$-anomaly, i.e., the boundary term  $b(\eta; A)$, which is an instance of classical higher anomaly. 
The DFM then automatically provides via \eqref{dressed-L} a higher cocyclic dressing $b(u; A)$,
\eqref{WZ-term-cocyclic-dressing}, which can then be understood as a higher Bardeen-Wess-Zumino counterterm cancelling the higher $\cT$-anomaly.

This facts strongly hints that our pursuit of the fully-fledged higher DFM framework developed within higher gauge theories -- e.g., as formalized within the language of higher connections \cite{Baez-Huerta2011} -- may find many clarifying applications, notably in supergravity and string/M-theory where higher form symmetries abound. 
In particular, the above observation suggests that the automatic \emph{quantum} anomaly cancellation mechanism built in the DFM, first described in \cite{JTF-Ravera2025-MFSRelaionalQuant, JTF-Ravera-InvQuantLetter2026}, extends naturally to  higher quantum anomalies, and could thus be expected to encompass Green-Schwarz-type mechanisms \cite{Alvarez-Gaume-Vazquez-Mozo2023, Green-Schwarz1984}.

\begin{acknowledgments}
J.F. is supported by the Austrian Science Fund (FWF), grant \mbox{[P 36542]},
and by the Czech Science Foundation (GA\v{C}R), grant GA24-10887S.
L.R. is supported by the project
GrIFOS, YOUR MSCA SoE,
CUP E13C24003600006, of which this work is part.
This study was supported by the European Union - NextGenerationEU, under the Italian National Recovery and Resilience Plan (PNRR), Mission 4 Component 2 Investment 1.2, funding scheme ``Young Researchers'' (D.D. 201 del 3.7.2024).
This manuscript reflects only the authors' views and opinions and the Ministry cannot be considered responsible for them.
\end{acknowledgments}

\section*{Data availability statement}

Data sharing not applicable to this article as no datasets were generated or analysed during the current study.

\section*{Author contribution declaration} 

Both authors share equal credit for the work done in this paper: J.F. and L.R. conceived the idea presented, J.F. and L.R. developed the formalism and did the calculations. Both J.F. and L.R. authors contributed equally to the writing of the manuscript. 

\bibliography{Biblio18}
\end{document}